\documentclass[amsmath,amssymb,pra,aps,showpacs,superscriptaddress,twocolumn]{revtex4-2}
\usepackage{amsmath,amsfonts,amssymb,amsthm,graphics,graphicx,epsfig,bbm}
\usepackage[colorlinks=true,citecolor=blue,linkcolor=blue,urlcolor=blue]{hyperref}
\usepackage[usenames]{color}
\usepackage{graphicx}
\usepackage{subfigure}
\usepackage{amsmath}
\usepackage{epsfig}
\usepackage{dcolumn}
\usepackage{bm}
\usepackage{color}
\usepackage{times}
\usepackage{epstopdf}
\usepackage{amssymb}
\usepackage{amstext}
\usepackage{latexsym}
\usepackage{hyperref}
\usepackage{amsfonts}
\usepackage{psfrag}
\usepackage{soul,xcolor}
\usepackage[normalem]{ulem}
\usepackage{dsfont}
\usepackage{txfonts}
\usepackage{float}

\newcommand{\ket}[1]{\vert #1 \rangle}
\newcommand{\bra}[1]{\langle #1 \vert}
\newcommand{\ketbra}[2]{\vert #1 \rangle \langle #2 \vert}
\newcommand{\braket}[2]{\langle #1 \vert #2 \rangle}

\newcommand{\norm}[1]{\| #1 \|}

\newcommand{\etal}{{\it{et al.}}}

\usepackage{tikz,xcolor,hyperref}
\definecolor{lime}{HTML}{A6CE39}
\DeclareRobustCommand{\orcidicon}{%
	\begin{tikzpicture}
	\draw[lime, fill=lime] (0,0) 
	circle [radius=0.16] 
	node[white] {{\fontfamily{qag}\selectfont \tiny ID}};
	\draw[white, fill=white] (-0.0625,0.095) 
	circle [radius=0.007];
	\end{tikzpicture}
	\hspace{-2mm}
}

\foreach \x in {A, ..., Z}{%
	\expandafter\xdef\csname orcid\x\endcsname{\noexpand\href{https://orcid.org/\csname orcidauthor\x\endcsname}{\noexpand\orcidicon}}
}

\begin{document}

\setstcolor{red}

\title{High-dimensional counterdiabatic quantum computing}
\date{\today}
\author{Diego Tancara \orcidB{}} 
\affiliation{Departamento de F\'isica, Universidad de Santiago de Chile (USACH), Avenida V\'ictor Jara 3493, 9170124, Santiago, Chile.}

\author{Francisco Albarr\'an-Arriagada \orcidA{}} 
\email[Corresponding author:]{\quad francisco.albarran@usach.cl}
\affiliation{Departamento de F\'isica, CEDENNA, Universidad de Santiago de Chile (USACH), Avenida V\'ictor Jara 3493, 9170124, Santiago, Chile.}

\begin{abstract}
    The digital version of adiabatic quantum computing enhanced by counterdiabatic driving, known as digitized counterdiabatic quantum computing, has emerged as a paradigm that opens the door to fast and low-depth algorithms. In this work, we explore the extension of this paradigm to high-dimensional systems. Specifically, we consider qutrits in the context of quadratic problems, obtaining the qutrit Hamiltonian codifications and the counterdiabatic drivings. Our findings show that qutrits can improve the solution quality up to 90 times compared to the qubit counterpart. We tested our proposal on 1000 random instances of the multiway number partitioning, max 3-cut, and portfolio optimization problems, demonstrating that, in general, without prior knowledge, it is better to use qutrits and, apparently, high-dimensional systems in general instead of qubits. Finally, considering the state-of-the-art quantum platforms, we show the experimental feasibility of our high-dimensional counterdiabatic quantum algorithms at least in a fully digital form. This work paves the way for the efficient codification of optimization problems in high-dimensional spaces and their efficient implementation using counterdiabatic quantum computing.
\end{abstract}

\maketitle

\section{Introduction}
Quantum computing has received significant attention in recent years. The emergence of quantum technology start-ups and increased funding from government agencies have led to unprecedented investment in the field. This surge is driven by the potential of quantum technologies to surpass the classical limits of traditional devices. Several experiments in quantum computing have claimed quantum advantage~\cite{AruteNature2019, ZhongPhysRevLett2021, WuPhysRevLett2021, MadsenNature2022, KimNature2023}, that is a computational problem where a quantum computer outperforms its classical counterpart. This breakthrough opens the door to solving previously intractable problems with potential industrial applications. 

Some notable areas of quantum computing application include quantum simulation, machine learning, and optimization problems~\cite{MalleyPhysRevX2016, BiamonteNature2017, Farhi2014}, where the paradigm of digital gate-based quantum computing takes advantage due to its universality~\cite{LloydScience1996, DiVincenzoFortschrPhys2000} and experimental implementations in different platforms~\cite{NakamuraNature1999, CiracPhysRevLett1995, KnillNature2001}. However, the current noisy intermediate-scale quantum (NISQ) devices present limitations, necessitating algorithmic improvements to achieve quantum advantage for industrial purposes~\cite{Preskill2018Quantum, Bharti2022RevModPhys}. 


Another interesting paradigm for quantum computing is adiabatic quantum computing (AQC), which provides promising algorithms to solve optimization problems~\cite{FarhiScience2001}. AQC involves an adiabatic evolution from a Hamiltonian with an easily generated ground state to a final Hamiltonian encoding the solution to the optimization problem. According to the quantum adiabatic theorem, the time-dependent state follows the instantaneous ground state during the adiabatic evolution. While AQC provides a standardized form to encode optimization problems in a Hamiltonian, its implementation is hampered by the long coherence times required for adiabatic evolution. One possible implementation of AQC can be done by digitalizing the adiabatic evolution~\cite{BarendsNature2016}. Nevertheless, due to the long evolution time required by adiabatic evolutions, the digitalization requests long circuit depth, which translates into low fidelity, being impractical for NISQ devices. It is essential to mention that the scalability of the total time of adiabatic evolution in AQC is equivalent to the scalability of the circuit depth for standard quantum computing algorithms~\cite{AharonovIEEE2004, Mizel2007PhysRevLett}.

To reduce the time of adiabatic protocols, shortcut to adiabaticity (STA) techniques emerge to accelerate these processes~\cite{OdelinRevModPhys2019}, specifically the counterdiabatic (CD) driving has yielded promising results \cite{DemirplakJPhysChemA2003, DemirplakJPhysChemA2005, Berry2009JPhysA}. However, the exact calculation of counterdiabatic fields is generally a complex task since it requires knowledge of the energy spectrum during evolution, which is impractical. In this line, approximate expressions for the CD terms open the door for implementable CD protocols~\cite{SelsPNAS2017, HatomuraPhysRevA2021, ClaeysPhysRevLett2019, MorawetzPhysRevB2024, CepaitePRXQuantum2023}. In this context, digitized counterdiabatic quantum algorithms (DCQA) have been recently proposed to enhance quantum algorithms on NISQ devices~\cite{Hegade2021PhysRevAppl}. This approach has been tested in various areas, including the improved quantum approximate optimization algorithm (QAOA)~\cite{WurtzQuantum2022, ChandaranaPhysRevRes2022}, as well as in factorization, biology, and finance problems~\cite{HegadePhysRevRes2022, HegadePhysRevA2021, ChandaranaPhysRevAppl2023}. 

Another possible enhancement for quantum computing is using high-dimensional quantum systems called qudits~\cite{LuoSciChinaPhysMechAstron2014}. The principal enhancement in using qudits is the amount of information stored compared to qubits, reducing the complexity of a quantum algorithm~\cite{WangFrontPhys2020} allowing more efficient codifications. This potential has led to the exploration of different physical platforms for the use of qudits; in this line, the generation of entangled qudits states has been theoretically proposed and experimentally implemented~\cite{PaesaniPhysRevLett2021, RingbauerNatPhys2022, CerveraRevAppl2022}, as well as, the use of qudits for efficient codifications in quantum simulations~\cite{VezvaeearXiv2024}.

In addition to these developments, quantum gates for high-dimensional quantum computing have been demonstrated across multiple platforms. For instance, single photons have been used to implement X-qudit gates~\cite{WangQuantumSciTechnol2021}, as well as, superconducting circuits and trapped ions have demonstrated the implementation of quantum gates for three-level (qutrits) and five-level systems~\cite{GossNatCommun2022, LuoPhysRevLett2023, HrmoNatCommun2023}. In this work, we study the application of the CD protocol for qutrits to address optimization problems and enhance algorithm performance compared to qubits. Specifically, we focus on the codification and performance of multiway number partitioning, max 3-cut, and portfolio optimization problems.

\section{Results}
\subsection{Hamiltonian codification}
We consider an evolution governed by a time-dependent Hamiltonian 

\begin{eqnarray}
    H(t) = (1-\lambda)H_{\textrm{I}} + \lambda H_{\textrm{F}}+\dot{\lambda} A_\lambda,
    \label{Eq01}
\end{eqnarray}
that is an adiabatic evolution boosted by counterdiabatic driving~\cite{Hegade2021PhysRevAppl}, where $H_{\textrm{I}}$ is the initial Hamiltonian, $H_{\textrm{F}}$ the final Hamiltonian and $A_\lambda$ is the adiabatic gauge potential which can be calculated using the nested-commutator expansion~\cite{ClaeysPhysRevLett2019}. $T$ is the total evolution time and $\lambda$ the schedule function. In general, $H_{\textrm{I}}$ considers only local terms, in which the ground state is a uniform superposition of all the states. In the qubit case $H_{\textrm{I}}=-\sum_j\sigma_j^x$. Following the same structure, we define the next initial Hamiltonian
\begin{equation}
    H_{\textrm{I}}=-\omega_0\sum_j X_j,
\end{equation}
where the subindex $j$ refers to the $j$th qutrit and the matrix $X$ is given by
\begin{eqnarray}
    X = \begin{pmatrix}
        1 & 1 & 0\\
        1 & 0 & 1\\
        0 & 1 & 1
    \end{pmatrix},
\end{eqnarray}
the ground state $\ket{\phi_g}$ of $H_{\textrm{I}}$ is given by 
\begin{eqnarray}
    \ket{\phi_g}=\bigotimes_{j}\ket{+_3}_j;\quad \ket{+_3}_j=\sqrt{\frac{1}{3}}(\ket{0}_j+\ket{1}_j+\ket{2}_j),
\end{eqnarray}
that is, the flat superposition of all the possible states in the computational basis for qutrits.

As we consider discrete classical optimization problems, the final Hamiltonian take the form:
\begin{equation}
    H_{\textrm{F}}=\sum_j\omega_j^{(1)}Z_j+\sum_{j,k}\omega_{j,k}^{(2)}Z_jZ_k + \sum_{j,k,l}\omega_{j,k,l}^{(3)}Z_jZ_kZ_l + \dots,
    \label{FinalHamiltonian}
\end{equation}
where we can consider all the $p$-local terms with
\begin{eqnarray}
    Z = \begin{pmatrix}
        1 & 0 & 0\\
        0 & 0 & 0\\
        0 & 0 & -1
    \end{pmatrix}.
\end{eqnarray}

It is important to mention that any discrete optimization problem can be mapped to a Hamiltonian of the form (\ref{FinalHamiltonian}) by rewritting the variables of the cost function in base three. In this work, we consider three paradigmatic examples: the multiway number partitioning problem, the max k-cut problem, and the portfolio optimization problem.

For multiway number partitioning problem we consider a set $S = \{ s_1, s_2, ... s_N\}$ of N numbers into three
partitions, obtaining the following Hamiltonian:

\begin{eqnarray}
\notag H_{\textrm{F}}
\notag =&& \frac{1}{4} \sum_{j,k}\Big(9G_jG_k-6G_jZ_k-12G_j\\
&& +4Z_j+5Z_jZ_k+4\Big)s_js_k,
\label{resultspartitioning}
\end{eqnarray}
where $s_i$ is a real number of the set $S$ and $G_j = Z_j^2$.

For max k-cut we consider a graph $\mathcal{G}(V,E)$ composed of $V$ vertices and $E$ edges. The idea is to classify the vertices in $k$ different groups to maximize the edges shared between the groups. Considering $k=3$ we obtain the following Hamiltonian:

\begin{eqnarray}
    H_{\textrm{F}}  
    =\sum_{(j,k)\in E}\left[(1+\alpha)G_jG_k + Z_jZ_k-\alpha(G_j+G_k)\right],
\end{eqnarray}
with $\alpha$ a penalty constant (see Methods for details).

In portfolio optimization the goal is to distribute a budget across assets to maximize expected returns while minimizing risk. Considering trinary representation for discrete variable description of portfolio optimization we obtain the following Hamiltonian:

\begin{eqnarray}
    \notag H_\textrm{F} &=&  
     \sum_{j=1}^n \sum_{k=1}^g \alpha_{jk} Q_{u} +\sum_{j,j'=1}^n \sum_{k,k'=1}^g\beta_{jj'}^{kk'}Q_{u}Q_{u'}+\theta_3 b^2,
\end{eqnarray}
where $Q_u = Z_u + I$ and the remaining parameters are constants determined by the problem formulation (see Methods for a complete calculation).

Finally, as we mention before, we consider an adiabatic gauge potential $A_{\lambda}$ given by the nested commutator expansion, obtaining $A_{\lambda}\sim[H_{\textrm{I}},H_{\textrm{F}}]$ (see Methods for details).

\subsection{Energy error: qutrits vs qubits}
To show the performance of our solution, we will consider how much our solution minimizes the energy of the final Hamiltonian. For this proposal, we use the metric defined by the percentual error of the energy at the end of the evolution given by
\begin{eqnarray}
    \mathcal{R} = \frac{E(T)-E_{g}}{\Delta E},
\end{eqnarray}
where $E(t) = \bra{\psi(t)} H(t/T) \ket{\psi(t)}$ is the mean instantaneous energy, and $E_{g}$ is the ground state energy of $H_{\textrm{F}}$. $\Delta E$ is the gap between the two different minimal energies of $H_{\textrm{F}}$. $\mathcal{R}$ told us how good our solution is related to the energy gap $\Delta E$. Also, we consider three different regimes: short-term, medium-time, and large-time regimes. First, we consider specific cases as examples of the performance of our high-dimensional counterdiabatic quantum computing (HDCQC) protocol compared to the standard qubit paradigm.
\begin{figure}[t]
    \centering
    \includegraphics[width=0.9\linewidth]{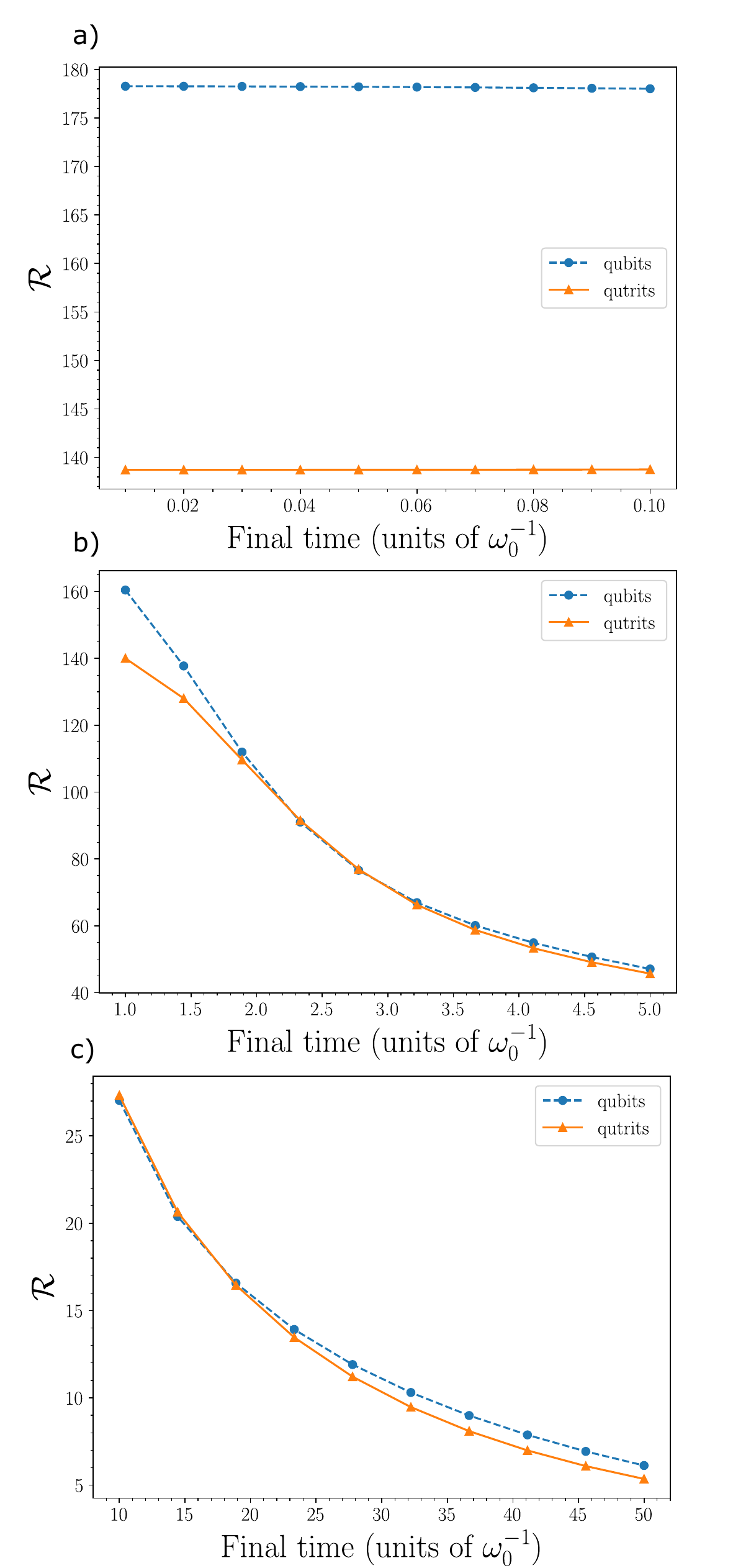}
    \caption{Three partitions multiway number partitioning problem for the set $\{0.8, 1.1, 1.1, 0.7, 1, 0.3\}$ for different regimes, that is, a: short final time, b: medium final time, and c: long final time.}
    \label{Figure01} 
\end{figure}

\begin{figure}[b]
\centering
\includegraphics[width=0.9\linewidth]{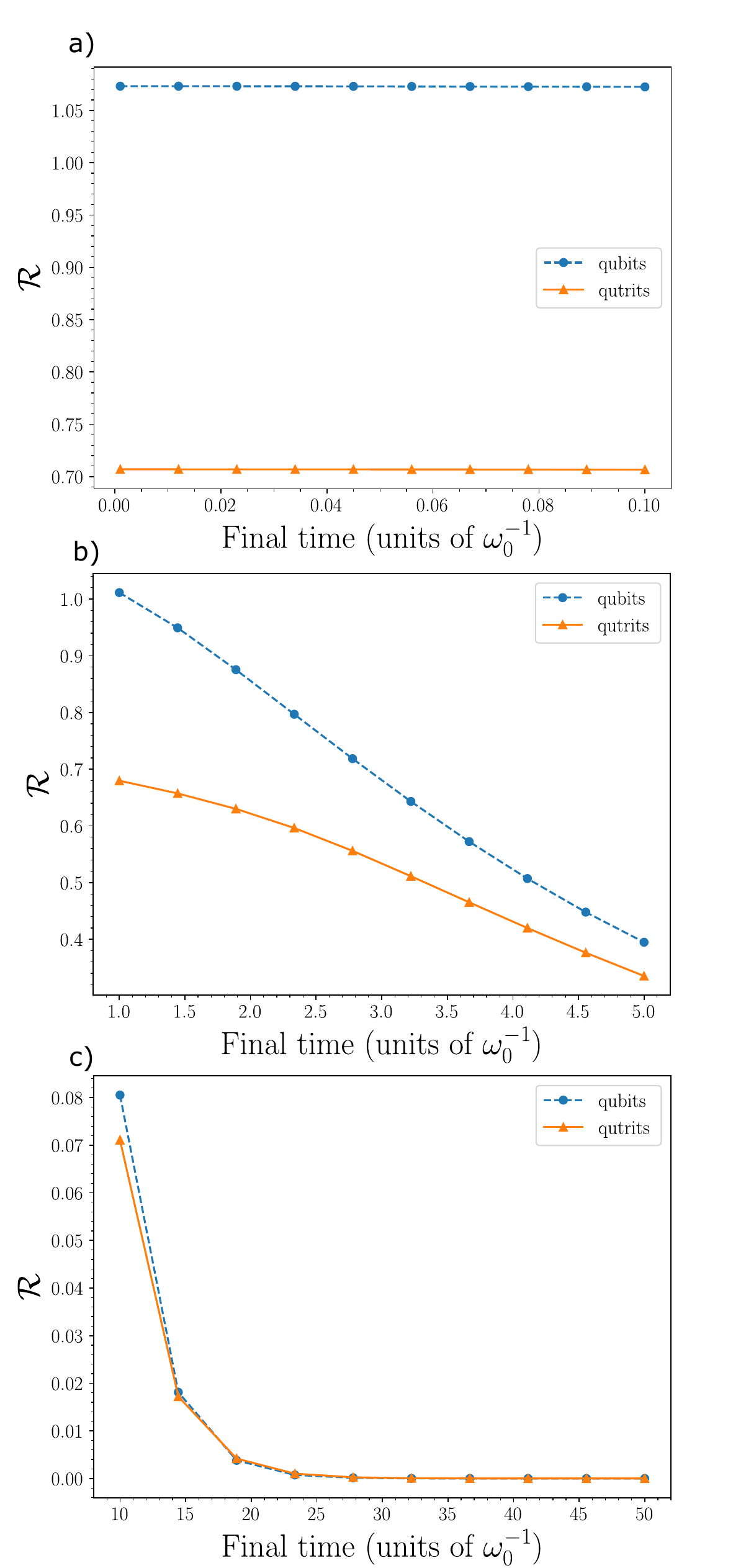}
\caption{Max k-cut for connectivity of wheels $W_6$ (6 nodes) for a: short final time, b: medium final time, and c: long final time.}   
\label{Figure02}
\end{figure}

Figure~\ref{Figure01} shows the $\mathcal{R}$ metric for the multiway number partitioning problem, where we consider the set $\{0.8, 1.1, 1.1, 0.7, 1, 0.3\}$. We observe a small advantage in using qutrits over qubits, which can be appreciated only in the impulse regime. We note that in Fig~\ref{Figure01} a, the value of $\mathcal{R}$ is high for both qutrit and qubits, which implies that even if we have a better performance for the qutrit case, it is far from the solution.

For the max 3-cut problem, Fig.~\ref{Figure02} shows the $\mathcal{R}$ metric for a six-nodes wheel graph ($W_6$). We can observe that in these cases, the performance of qutrits is appreciable even in the middle time regime, as can be seen in Fig.~\ref{Figure02} a and b. Also, the values for $\mathcal{R}$ go below $1$ in all the regimes, which indicates that our solutions are close to the ground state even for the impulse regime. 

\begin{figure}[t]
    \centering
    \includegraphics[width=0.9\linewidth]{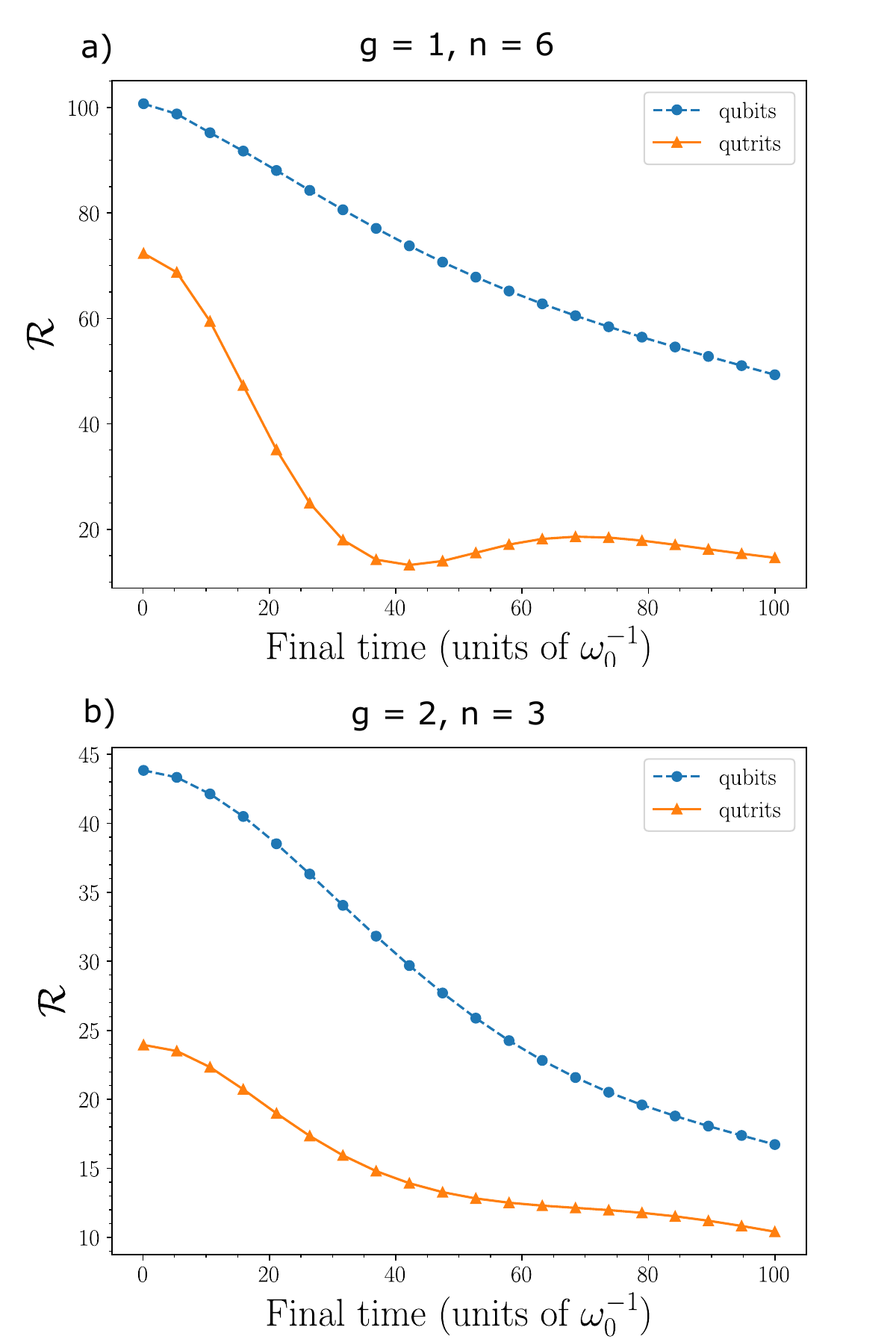}
    \caption{Portfolio optimization, comparing qubits and qutrits for a: $g=1$, $n=6$ and b: $g=2$, $n=3$.}\label{Figure03}
    \end{figure}

Fig.~\ref{Figure03} shows the performance using qutrits and qubits for the portfolio optimization problem. For example, we consider the data provided by \textit{yfinance} \cite{Ran2023yfinance} for the assets: Apple, Microsoft, Google, Amazon, Tesla, and Netflix for the period 2020-2023. Using this data, the mean expected returns and covariance matrix are

\begin{eqnarray}
    m &=& 10^{-4} (10.24, 1.411, 5.730, 8.082, 4.122, 29.64)\nonumber\\
    \rho &=& 10^{-4} \begin{pmatrix}
    5.413 &  3.799 & 3.696 & 4.133 &3.690 & 5.495\\
 3.799 & 6.062 & 3.638 & 3.781 & 4.768 & 5.343\\
 3.696 & 3.638 & 4.730 & 3.953 & 3.543 & 4.474\\
 4.133 & 3.781 & 3.953 & 4.794 & 3.631 & 4,987\\
 3.690 & 4.77 & 3.543 & 3.631 & 1.067 & 6.067\\
 5.495 & 5.343 & 4.474 & 4.987 & 6.067 & 20.67    
    \end{pmatrix}.
\end{eqnarray}

For Fig.~\ref{Figure03} a, we consider the mentioned six assets, observing that the improvement of qutrits over qubits is appreciable over all the evolution. In the mentioned figure, we use $g=1$, which means $G_f=1/2$. Next, in Fig.~\ref{Figure03} b, we consider only three first assets of the above list, but using g=2, which means that $G_f=1/8$. We can see in this figure that again, the performance of qutrits over qubits is appreciably better for the range $T\in [0,100]$, nevertheless in all the cases $\mathcal{R}>10$, which suggests that even for $T=100\omega_0^{-1}$ we are not reaching the adiabatic regime.
    
\subsection{Sucess probability enhancement}
To understand the enhancement produced by the use of qutrits instead of qubits, we define the success probability using $j$- dimensional systems as:

\begin{eqnarray}
    P_j(T) = \sum_{g=1}^{d_{j}} |\braket{g_j | \psi_j(T)}|^2,
\end{eqnarray}
where are $d$ degenerate groundstates, and $j$ is the dimension of the qudits used in the computational process, $j=2$ for qubits, and $j=3$ for qutrits. The time $T$ represents the total time of the evolution, then $\lim_{T\rightarrow\infty}P_j(T)$=1. To obtain a metric for the enhancement, we can define the success probability enhancement for a time $T$ as:
\begin{eqnarray}
    \mathcal{P}_{j/k}(T) = \frac{P_{j}(T)}{P_{k}(T)},
\end{eqnarray}
in our case we will focus in $\mathcal{P}_{3/2}(T)$, that is the comparison between qutrits and qubits. Also, we must remember that we are considering evolutions with the CD term given by the approximation for the AGP in Eq.~(\ref{Eq26}). To obtain representative results, we consider the success probability enhancement for 1000 random instances for each problem described in the previous section, and again for different total times $T=0.1\omega_0^{-1}$, $T=\omega_0^{-1}$ and $T=10\omega_0^{-1}$. 

\begin{figure}[t]
    \centering
    \includegraphics[width=0.9\linewidth]{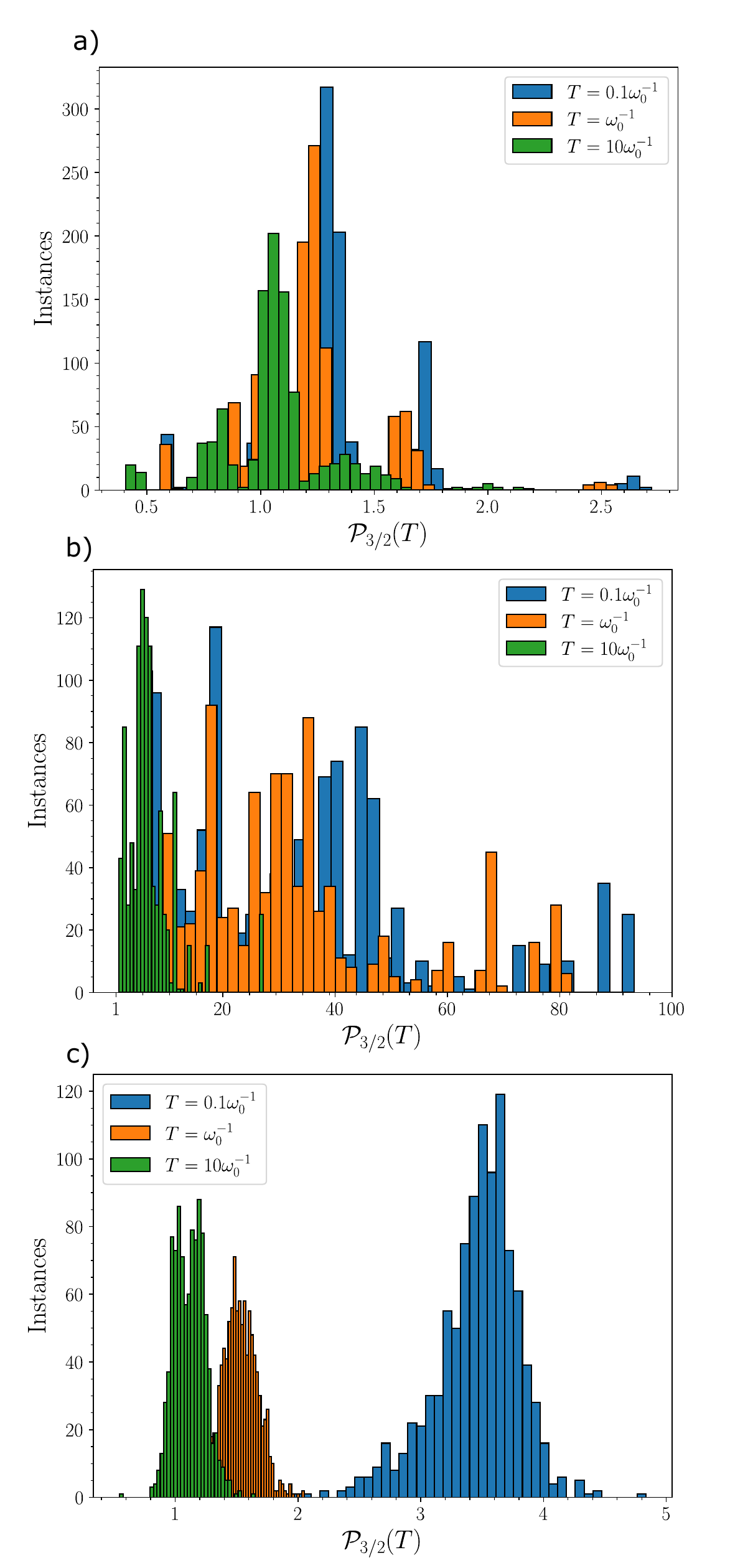}
    \caption{Success probability enhancement for 1000 random instances for a: multiway number partitioning for six random numbers in the range $[0,1]$, b: Max 3-cut for six nodes graphs with random connectivity, and c: Portfolio optimization for six assets with random expected return and covariance matrix.}   
    \label{Figure04}
    \end{figure}

Figure~\ref{Figure04} a shows the results of $\mathcal{P}_{3/2}(T)$ for the multiway number partitioning problem for a set of six random numbers in the range $[0,1]$, where we can see that more considerable enhancement can be reached for short time $T$ as we can expect. In this case, we can observe that there are instances for which $\mathcal{P}_{3/2}(T)<1$, that is that the use of qutrits does not provide a benefit, but in most cases the obtain $\mathcal{P}_{3/2}(T)>1$. 

Figure~\ref{Figure04} b shows the results of $\mathcal{P}_{3/2}(T)$ for the max 3-cut problem for random graphs of six nodes, where we can see that more considerable enhancement can be reached for short time $T$ as we can expect. In this case, we can observe that for all the 1000 random instances $\mathcal{P}_{3/2}(T)>1$, which suggests that for this problem always is better the use of qutrits instead of qubits. Also, it is necessary to remark that there are instances where $\mathcal{P}_{3/2}(T)>90$ which indicates that the use of the qutrits can enhance the solution around two orders of magnitudes.

It is interesting to note that according to our results, there are cases where the use of qutrits is not better for qubits, even if the codification of the problem is more natural for qutrits. Nevertheless, if we consider the mean success probability enhancement ($\bar{\mathcal{P}}_{3/2}$), we can note that the cases under study always is larger than $1$. For example for the multiway number partitioning problem we have $\bar{\mathcal{P}}_{3/2}(0.1\omega_0^{-1})=1.31$, $\bar{\mathcal{P}}_{3/2}(\omega_0^{-1})=1.23$ and $\bar{\mathcal{P}}_{3/2}(10\omega_0^{-1})=1.07$, and the percentage of instances with enhancement larger than 1 are $84.6\%$, $79.8\%$ and $74.8\%$ for the same times respectively. It suggests that without previous knowledge, it is a good idea to codify and use qutrits instead of qubits. 

On the other hand, for the max 3-cut problem in all the cases, we obtain enhancement, where the mean success probabilities enhancement are $\bar{\mathcal{P}}_{3/2}(0.1\omega_0)=35.34$, $\bar{\mathcal{P}}_{3/2}(\omega_0)=30.41$ and $\bar{\mathcal{P}}_{3/2}(10\omega_0)=6.99$, which indicates that an efficient codification in high dimensional systems always is better for max 3-cut problem.

For the portfolio optimization problem we have that for $T=0.1\omega_0^{-1}$ and $T=1\omega_0^{-1}$ all the cases get enhancement, where the mean success probability enhancement are $\bar{\mathcal{P}}_{3/2}(0.1\omega_0)=3.56$ and $\bar{\mathcal{P}}_{3/2}(\omega_0)=1.54$. For $T=10\omega_0^{-1}$, we have that the percentage of instances with enhancement surpass $80\%$ with a mean enhancement $\bar{\mathcal{P}}_{3/2}(10\omega_0)=1.12$, which indicates that an efficient codification in high dimensional systems provides in general enhancement for portfolio optimization, less than in the max 3-cut problem but more that in the multiway number partitioning problem.
\begin{figure}[t]
    \centering
    \includegraphics[width=0.8\linewidth]{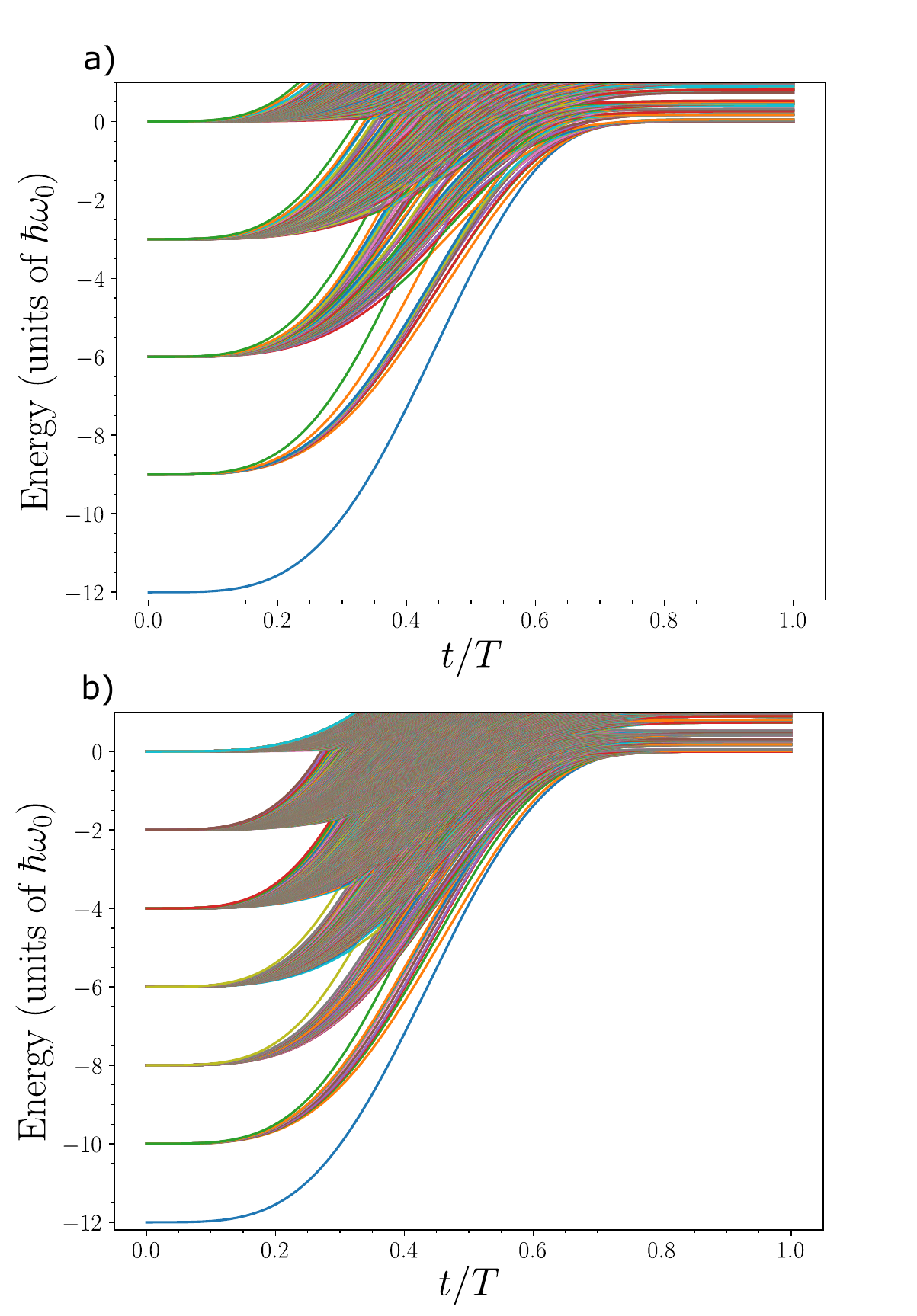}
    \caption{Spectrum of adiabatic Hamiltonian for multiway number partitioning problem for the set  $[8/10, 11/10, 15/10, 7/10, 10/10, 3/10]$. a: Qutrits implementation and b: qubits implementation.}
    \label{Figure05}
    \end{figure}
\begin{figure}[t]
    \centering
    \includegraphics[width=0.8\linewidth]{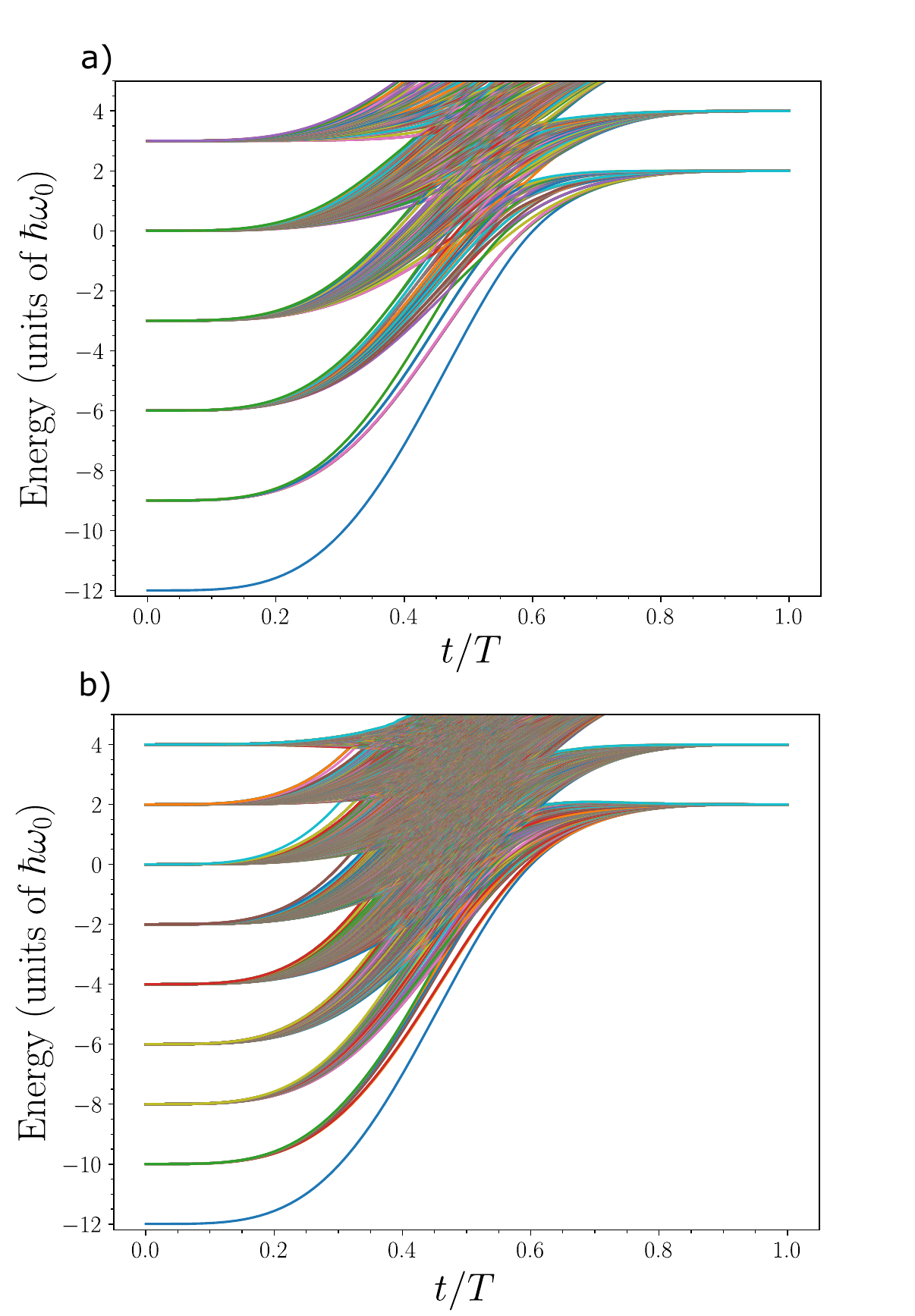}
    \caption{Spectrum of adiabatic Hamiltonian for max 3-cut problem with $W_6$ connectivity. a: Qutrits implementation and b: qubits implementation.}
    \label{Figure06}
    \end{figure}

\begin{figure}[t]
    \centering
    \includegraphics[width=0.8\linewidth]{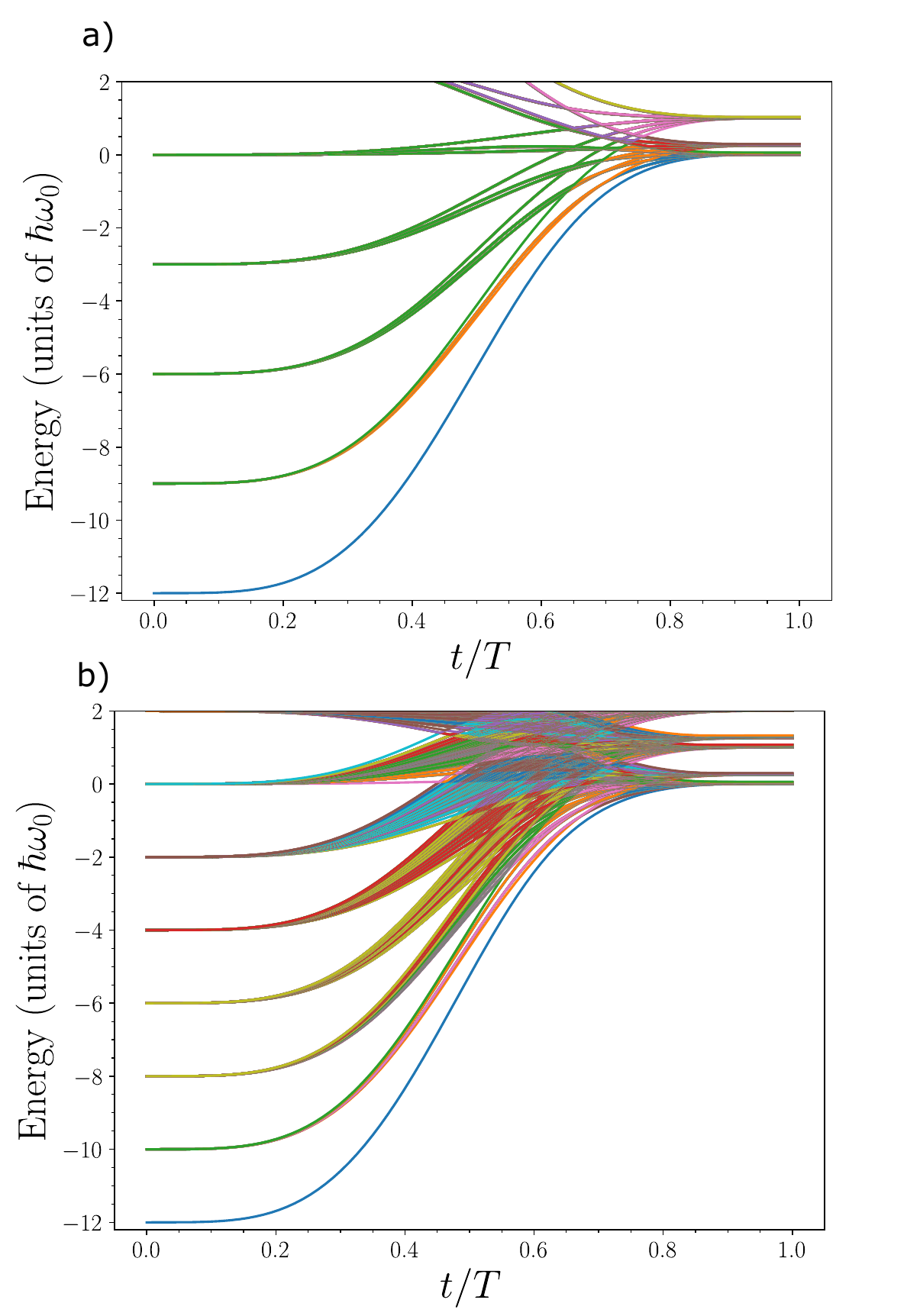}
    \caption{Spectrum of adiabatic Hamiltonian for portfolio optimization problem with $g=1$ and six assets ('AAPL', 'MSFT', 'GOOGL', 'AMZN', 'TSLA', 'NFLX') a: Qutrits implementation and b: qubits implementation.}
    \label{Figure07}
    \end{figure}    

\begin{figure}[b]
    \centering
    \includegraphics[width=0.8\linewidth]{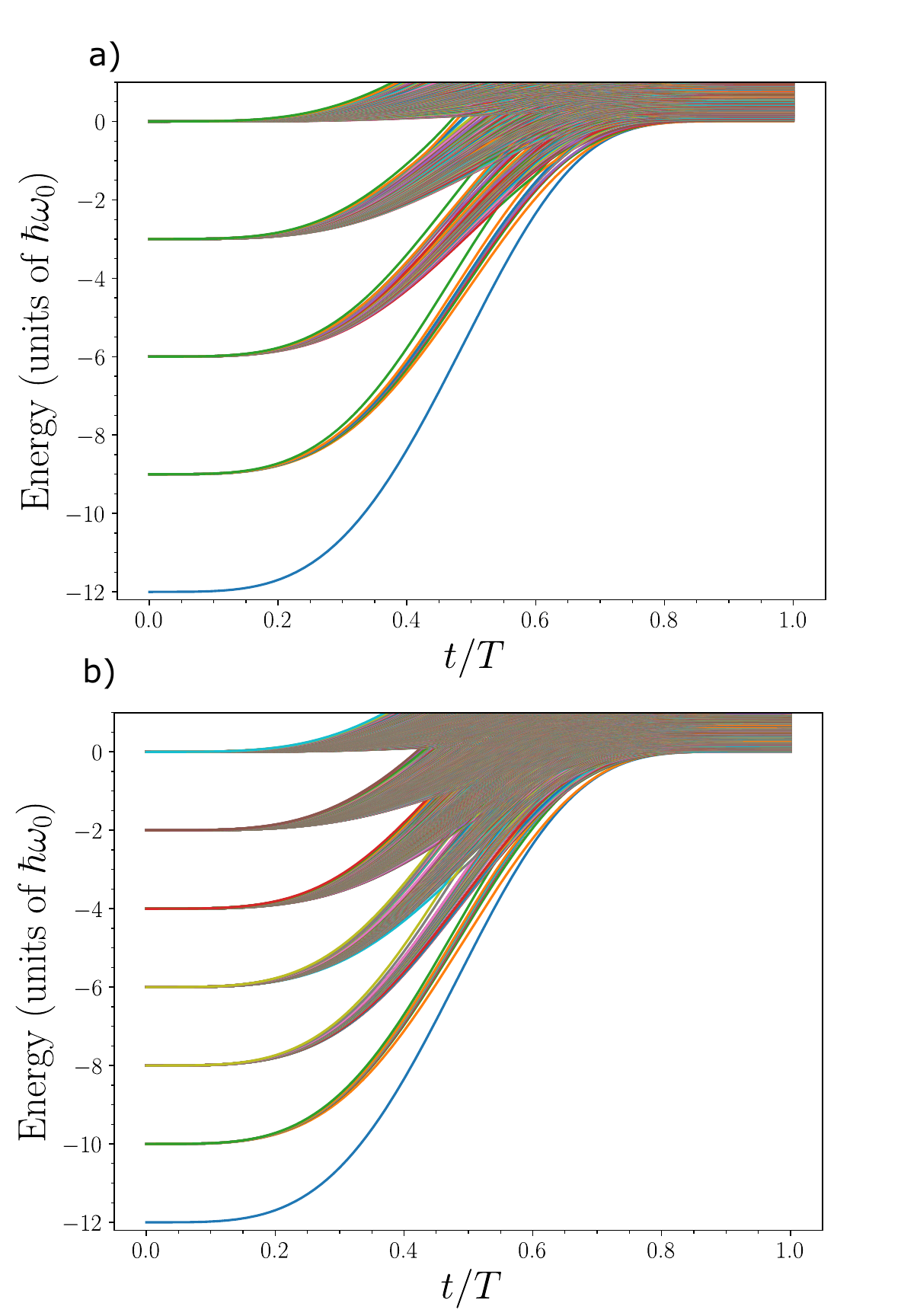}
    \caption{Spectrum of adiabatic Hamiltonian for multiway number partitioning problem for the set: $[0.252, 0.205, 0.123, 0.798, 0.726, 0.086]$, that has $\mathcal{P}_{3/2}=0.56$. a: Qutrits implementation and b: qubits implementation.}
    \label{Figure08}
    \end{figure}

\section{Discussion}
To understand why the enhancement in the performance of qutrits depended strongly on each case, we plot the energy spectrum of the adiabatic Hamiltonian during a time-evolution in Figs.~\ref{Figure05},\ref{Figure06}, and \ref{Figure07} to see the behavior of it for different cases. A small energy gap between the ground state and the first excited energy at the final time, combined with a higher density of levels throughout the evolution, increases the chances of level crossings and the risk that the system ends up in an excited state rather than in the ground state at the end of the evolution. In our best case, the max 3-cut problem (Fig. \ref{Figure06}), the energies of the possible solutions (degenerate ground state) are far from the following energy, obtaining few level crossings. Now, even if, for the portfolio optimization problem (Fig. \ref{Figure07}), the final energy gap is smaller than in the number partitioning problem (Fig. \ref{Figure05}), we have more level crossings in the last one, which favors the unwanted energy transitions. 

In general, we observe a clear difference between the qubit and qutrit cases, where a higher density of levels near the ground state during the evolution appears in the qubit encoding. Therefore
in the specific cases under study, qutrit, in general, reduces the number of crossing levels, resulting from a more efficient codification that avoids introducing useless redundancy. An opposite case is the multiway number partitioning problem for the set $[0.252, 0.205, 0.123, 0.798, 0.726, 0.086]$, which spectrum in time is given in Fig. \ref{Figure08}, and $\mathcal{P}_{3/2}=0.56$. In the mentioned figure, we can observe that in the middle of the evolution, it is hard to distinguish which case has less level crossing, reinforcing the idea that the performance of our HDCQC is related to the density of level crossings presented during the evolution.

In conclusion, we study the use of high-dimensional systems (qutrits) for quantum computing in the counterdiabatic quantum computing paradigm. Specifically, we study the performance of a qutrit codification for the multiway number partitioning, max 3-cut, and portfolio optimization problem. We find that, in general, the use of qutrits provides better results than the qubits counterpart, where in some cases the enhancement factor reaches $90$, that is, almost two magnitudes orders, where the enhancement is related to the density of level crossing during the evolution, being the case with more enhancement the max 3-cut problem, with a mean enhancement of 35.34, 30.41 and 6.99 for final time $T=\{0.1\omega_0^{-1}, \omega_0^{-1}, 10\omega_0^{-1}\}$, even if, for the worst cases, that is number partitioning problem we have that the mean enhancement are 1.31, 1.23 and 1.07 for the same final times, suggesting that the use qutrit and possible high-dimensional system is better for an efficient codification. 

Finally, we analyze the current state-of-the-art in trapped ions, which allows the experimental implementation of universal qutrit quantum computing, allowing the implementation of the digital version of our high-dimensional counterdiabatic quantum computing proposal. This work provides numerical evidence of the benefit of using high-dimensional systems, specifically qutrits instead qubits. It shows its feasibility according to the current state-of-the-art technology, providing evidence suggesting that the use of qutrits opens a potential path for efficient codifications and therefore better performance of current algorithms, where the extension to high-dimensional systems beyond qubits looks promising.

\section{Methods}
We are focused on an adiabatic evolution enhanced by counterdiabatic drivings for high-dimensional systems, specifically for qutrits. In this section we describe the framework to obtain the initial and final Hamiltonian for different problems, we briefly introduce the nested commutator approach to get an approximation for the counterdiabatic driving.


\subsection{Multiway number partitioning problem}

This problem considers a set $S$ of $N$ numbers into different partitions. The goal is that the sum of the elements of each partition be as equal as possible to the others. We will consider the specific case of three partitions and then use qutrits to codify them. The set $S$ is defined as
\begin{eqnarray}
    S = \{ s_1, s_2, ... s_N\},
\end{eqnarray}
where $s_j$ is a real number. As usual, we consider three different labels, one for each partition. Each label is a value of a \textit{trinary} variable; we consider as many trinary variables as numbers in the set $S$. Specifically, we use the trinary variable $q_j=\{1,0,-1\}$, where $q_j=1$ is $s_j$ belongs to the first partition, $q_j=0$ is $s_j$ is in the second partition and $q_j=-1$ if $s_j$ is an element of the third partition. Now, we define $C_k$ the sum of the elements in the partition $k$, then we have that:
\begin{eqnarray}
    C_1 &=& \sum_{j=1}^N\frac{1}{2}q_j\left(q_j+1\right)s_j\\
    C_2 &=& \sum_{j=1}^N\left(1-q_j\right)\left(1+q_j\right)s_j\\
    C_3 &=& \sum_{j=1}^N\frac{1}{2}q_j\left(q_j-1\right)s_j.\label{sums}
\end{eqnarray}

As in this case, the goal is that $C_1$, $C_2$ and $C_3$ be as equal as possible; the minimization is over the function
\begin{equation}
    F=\left(C_1-C_2\right)^2+\left(C_1-C_3\right)^2.
\end{equation}

We note that the term $(C_2-C_3)$ can be neglected by transitivity. Changing the trinary variables $q_j$ to qutrits operators $Z_j$ we obtain the next Hamiltonian
\begin{eqnarray}
\notag H_{\textrm{F}}=&&\frac{1}{4}\left[\sum_j\left(3G_j-Z_j-2\right)s_j\right]^2 + \left[\sum_jZ_js_j\right]^2\\
\notag =&& \frac{1}{4} \sum_{j,k}\Big(9G_jG_k-6G_jZ_k-12G_j\\
&& +4Z_j+5Z_jZ_k+4\Big)s_js_k,
\label{partitioning}
\end{eqnarray}
where $G_j=Z_j^2$ is a one-body operator of the form 
\begin{equation}
    G_j= \begin{pmatrix}
        1 & 0 & 0\\
        0 & 0 & 0\\
        0 & 0 & 1
    \end{pmatrix}.
\end{equation}
Therefore, the Hamiltonian (\ref{partitioning}) has only one and two body operators corresponding to a quadratic form.

\subsection{Max k-cut}

In this problem, we consider a graph $\mathcal{G}(V,E)$ composed of $V$ vertices and $E$ edges. The idea is to classify the vertices in $k$ different groups to maximize the edges shared between the groups. We consider the max 3-cut problem, meaning we consider three partitions. Similar to the previous case, we use a trinary variable for each vertex to indicate which partition it belongs to. As each variable $q_j$ represents the label of the vertex $v_j$, we need to minimize the case when two vertices come from different positions. If we consider the product $q_j q_k$, it is $1$ if $q_j=q_k$ (except for $q_j=0$), and $0$ and $-1$ in other case. It means that the minimal value of the term $q_jq_k(q_jq_k+1)$ is obtained for $q_j\neq q_k$ (except for $q_j=q_k=0$). 

Therefore, our minimization function is the summation of the above term for the vertices $j$ and $k$ that share an edge, and for the case $q_j=q_k=0$ we add a penalization term obtaining
\begin{eqnarray}
    F = \sum_{(j,k)\in E}\left[q_jq_k\left(q_jq_k+1\right)+\alpha\left(1-q_j^2\right)\left(1-q_k^2\right)\right],
\end{eqnarray}
where $\alpha$ is the penalty constant. Transforming the trinary variable to qutrit operators, we obtain the following Hamiltonian for max 3-cut:
\begin{eqnarray}
    H_{\textrm{F}} = && \sum_{(j,k)\in E}\left[Z_j Z_k\left(Z_jZ_k+1\right)+\alpha\left(1-Z_j^2\right)\left(1-Z_k^2\right)\right]\nonumber\\
    =&&\sum_{(j,k)\in E}\left[(1+\alpha)G_jG_k + Z_jZ_k-\alpha(G_j+G_k)\right],
\end{eqnarray}
where we obtained a quadratic Hamiltonian, as in the previous case.

\subsection{Portfolio Optimization}
The last example we consider is the Markowitz portfolio optimization~\cite{MarkowitzJFinance1952}. The goal is to distribute a budget $b$ across $n$ assets to maximize expected returns while minimizing risk. To estimate the expected return, we consider the historical market return data, while the risk is quantified using the covariance matrix of the historical data. The cost function for this problem is given by:

\begin{equation}
    F = \theta_1\sum_{j=1}^n m_j p_j+\theta_2\sum_{j,j'=1}^n \rho_{j,k}p_jp_j'+\theta_3\left(\sum_{j=1}^nG_f b p_j-b\right)^2,
\end{equation}
where $m_j$ is the expected return of asset $j$, $\rho_{j,k}$ is the covariance matrix element that shows the variance of the $k$th asset with the $j$th asset, and $p_j$ is an integer related to the portion of the $j$th asset in the portfolio. $G_f$ is the granularity function such that $\sum_j G_fp_j=1$, that is a normalization constant. For example, if we consider $g$ trinary digits per assets $G_f = 1/(3^g-1)$. Additionally, $\theta_1, \theta_2, \theta_3$ are the Lagrange multipliers, pondering the expected returns, risk, and budget constraints, respectively.

In a trinary representation, we gave
\begin{eqnarray}
    p_j = \sum_{k=1}^g3^{k-1}q_{u_{j,k}},
\end{eqnarray}
where the index $u_{j,k}=(j-1)g+k$, and $q_{j}\in \{0,1,2\}$, the index $j$ is the label for the asset and $k$ the term in the sumatory. From now on, we will refer to $u_{j,k}$ as $u$. As we consider $g$ trinary digits per asset, we need to consider $ng$ trinary digits for the total problem.

To obtain the Hamiltonian, we replace the trinary variables by
\begin{eqnarray}
    q_{u}  \rightarrow Q_u = Z_{u}+I,
\end{eqnarray}
in the cost function, where $I$ is the identity for qutrits and $Q_u$ is one body operator of the form:
\begin{eqnarray}
    Q_u = \begin{pmatrix}
        2 & 0 & 0 \\
        0 & 1 & 0\\
        0 & 0 & 0       
    \end{pmatrix}.
\end{eqnarray}
Therefore, in terms of $Q_u$, the Hamiltonian has the following form
\begin{eqnarray}
    \notag H_\textrm{F} &=&  \theta_1\sum_{j=1}^n m_j \sum_{k=1}^g3^{k-1}Q_{u} \\
    \notag&+& \theta_2\sum_{j,j'=1}^n \rho_{j,j'}\sum_{k,k'=1}^g3^{k-1}Q_{u}3^{k'-1}Q_{u'} \\ \notag &+&\theta_3\left(\sum_{j=1}^nG_f b \sum_{k=1}^g3^{k-1}Q_{u}-b\right)^2\\
    &=& \sum_{j=1}^n \sum_{k=1}^g \alpha_{jk} Q_{u} +\sum_{j,j'=1}^n \sum_{k,k'=1}^g\beta_{jj'}^{kk'}Q_{u}Q_{u'}+\theta_3 b^2,
\end{eqnarray}
where:
\begin{eqnarray}
    \notag \alpha_{jk} = 3^{k-1}(\theta_1 m_j - \theta_3 2b^2 G_f),\\
    \beta_{jj'}^{kk'} = 3^{k+k'-2}(\theta_2\rho_{jj'}+\theta_3 G_f^2 b^2).
\end{eqnarray}
Following the codification strategies for the multiway number partitioning and max 3-Cut problems, the portfolio optimization Hamiltonian is constructed analogously by mapping the relevant trinary variables into qutrit operators but with the $Q_u$ operator. This codification preserves the cost function structure and can be expressed in terms of local and bilocal qutrit operators, preserving a physically implementable structure suitable for counterdiabatic quantum computing. Furthermore, this framework is flexible and can be adapted to a broader class of optimization problems beyond those considered in this work.
\subsection{Counterdiabatic drivings}
\label{CD_section}
To accelerate adiabatic protocols, we can use counterdiabatic drivings, a control field to suppress unwanted transitions between instantaneous eigenstates. The time-dependent Hamiltonian modified by the addition of the CD term reads
\begin{eqnarray}
    H(t) = \underbrace{(1-\lambda)H_{\textrm{I}} + \lambda H_{\textrm{F}}}_{H_{\textrm{ad}}}+\dot{\lambda} A_\lambda,
    \label{Eq20}
\end{eqnarray}
where we use the following schedule function:

\begin{equation} 
\lambda(t) = \sin^2 \left ( \frac{\pi}{2} \sin^2 \left (\frac{\pi t}{2 T} \right ) \right ), \label{Eq09}
\end{equation} 
and $A_\lambda$ is called the adiabatic gauge potential (AGP), where the time dependence is given in the schedule function $\lambda$. The AGP satisfy
\begin{eqnarray}
    [i\partial_\lambda H_{\textrm{ad}} - [A_\lambda, H_{\textrm{ad}}], \hat{H}_{\textrm{ad}}]=0.
\end{eqnarray}
The exact form of $A_\lambda$ is determined by the instantaneous eigenstates of $H_{\textrm{ad}}$, being impractical to calculate~\cite{Berry2009JPhysA}. Nevertheless, there are approximations to the ADG such that:
\begin{eqnarray}
    A_\lambda^{l} = i\sum_{k=1}^{l}\alpha_k O_{2k-1},
\end{eqnarray}
with 
\begin{eqnarray}
    O_k = \underbrace{[H_{\textrm{ad}},[H_{\textrm{ad}},...[H_{\textrm{ad}}}_{k},\partial_\lambda H_{\textrm{ad}}]]].
\end{eqnarray}
The coefficients $\alpha_k$ satisfy the next linear system of equations~\cite{XiePhysRevB2022}:
\begin{eqnarray}
    \begin{pmatrix}
\Gamma_2 & \Gamma_3 & \cdots & \Gamma_{l+1} \\
\Gamma_3 & \Gamma_4 & \cdots & \Gamma_{l+2} \\
\vdots & \vdots & \ddots & \vdots \\
\Gamma_{l+1} & \Gamma_{l+2} & \cdots & \Gamma_{2l}
\end{pmatrix}
\begin{pmatrix}
\alpha_1 \\
\alpha_2 \\
\vdots \\
\alpha_l
\end{pmatrix}
=
-\begin{pmatrix}
\Gamma_1 \\
\Gamma_2 \\
\vdots \\
\Gamma_l
\end{pmatrix},\label{alphas}
\end{eqnarray}
where $\Gamma_k = \norm{\hat{O}_k}^2_F$ is the squared Frobenius norm of the nested commutator. For $l=1$ we have
\begin{equation}
    \alpha_1=-\frac{\Gamma_1}{\Gamma_2}.
\end{equation}

In this work, we will focus on $l=1$; then, we will consider the next approximation for the AGP
\begin{equation}
    A_{\lambda}\approx i \frac{\Gamma_1}{\Gamma_2}[H_{\textrm{ad}},\partial_{\lambda}H_{\textrm{ad}}]=i \frac{\Gamma_1}{\Gamma_2}[H_{\textrm{I}},H_{\textrm{F}}]
    \label{Eq26}
\end{equation}.

\subsection{Physical implementation}
One possible implementation of high-dimensional counterdiabatic quantum computing (HDCQC) is by the digitalization of the adiabatic evolution enhanced with counterdiabatic drivings as proposed for qubits in ref~\cite{Hegade2021PhysRevAppl} and then implement it in a quantum platform that allows universal quantum computing with qutrits.

The basic idea is to write the time evolution governed by the Hamiltonian (\ref{Eq20}) as
\begin{eqnarray}
    U(t)&=&\mathcal{T}e^{\frac{-i}{\hbar}\int_0^{t}\left\{[1-\lambda(\tau)]H_{\textrm{I}}+\lambda(\tau)H_{\textrm{F}}+\dot{\lambda}(\tau)A_{\lambda}(\tau)\right\}d\tau}\nonumber\\
    &\approx&e^{\frac{-i}{\hbar}\sum_j\left\{[1-\lambda(j\Delta t)]H_{\textrm{I}}+\lambda(j\Delta t)H_{\textrm{F}}+\dot{\lambda}(j\Delta t)A_{\lambda}(j\Delta t)\right\}\Delta t}\nonumber\\
    &\approx&\prod_j e^{\frac{-i}{\hbar}\left\{[1-\lambda(j\Delta t)]H_{\textrm{I}}+\lambda(j\Delta t)H_{\textrm{F}}+\dot{\lambda}(j\Delta t)A_{\lambda}(j\Delta t)\right\}\Delta t}\nonumber\\
    &\approx&\prod_j e^{\frac{-i}{\hbar}[1-\lambda(t_j)]H_{\textrm{I}}\Delta t}e^{\frac{-i}{\hbar}\lambda(t_j)H_{\textrm{F}}\Delta t}e^{\frac{-i}{\hbar}\dot{\lambda}(t_j)A_{\lambda}(t_j)\Delta t}, \label{eq31}
\end{eqnarray}
where $t_j=t_{j-1}+\Delta t=j\Delta t$. If $\Delta t$ is small enough this approximation reproduce the adiabatic evolution enhanced by counterdiabatic drivings.

Now, an arbitrary unitary transformation in high-dimensional systems can be implemented by the suitable combination of all two-level rotations of the high-dimensional system plus a two-body controlled phase gate. Such transformations have been reported in different physical platforms, where superconducting circuits and trapped ions are promising for universal qutrit quantum computing.
For example, in Ref.~\cite{KilmovPhysRevA2003} a two-level qutrit rotation has been proposed using Zeeman's level structure in trapped ions, where they consider a model with dipole and rotating wave approximations and five levels $i=0,1,2,3,4$, with two of them ($3,4$) are adiabatically eliminated. For this case, considering only carrier transitions and detuning conditions, we obtain the following evolution operator:
\begin{equation}
    U_{\phi,k,k'}= \begin{pmatrix}
        1+|g(k)|^2 C(\phi) & g(k) g^{*}(k')C(\phi) & -ig(k) \sin(\phi) \\
        g(k') g^{*}(k) C(\phi)& 1+|g(k)|^2 C(\phi)  & -ig(k) \sin(\phi)\\
        -ig^{*}(k) \sin(\phi) & -ig^{*}(k') \sin(\phi) & \cos(\phi) 
    \end{pmatrix} \label{Utrappedions}\, 
\end{equation}
where $\phi = \Omega t$, $\Omega = |k|^2+ |k'|^2 $ with $k = \Omega_{40}\Omega_{42}^{*}/\Delta$ and $k' = \Omega_{30}\Omega_{31}^{*}/\Delta$ where $\Omega_{i,j}$ are coupling between level $i$ and $j$, $g(k) = k/\Omega$ and $C(\phi) = \cos(\phi)-1$. By setting $\phi$, $k$, and $k'$ in Eq.~(\ref{Utrappedions}), we can implement any qutrit rotation, for example, two-level transformations of the form \cite{KilmovPhysRevA2003}:
\begin{eqnarray}
    \notag U_{ij} = \cos(\alpha_{ij})(\ket{i}\bra{i}+\ket{j}\bra{j})\\-e^{i\beta_{ij}}\sin(\alpha_{ij})\ket{i}\bra{j}-e^{-i\beta_{ij}}\sin(\alpha_{ij})\ket{j}\bra{i},
\end{eqnarray}
can be obtained by fixing
\begin{eqnarray}
    \cos(\alpha_{ij}) =
\begin{cases}
\frac{|k|^2-|k'|^2}{|k|^2+|k'|^2} & \text{if } i=1, j=2 \\ \cos(|k|t) & \text{if } i =0, j=2\\
\cos(|k'|t) & \text{if } i =0, j=1
\end{cases},
\end{eqnarray}

\begin{eqnarray}
    e^{i\beta} =
\begin{cases}
\frac{|k|^2-|k'|^2}{|k|^2+|k'|^2} & \text{if } i=1, j=2 \\ \cos(|k|t) & \text{if } i =0, j=2\\
\cos(|k'|t) & \text{if } i =0, j=1
\end{cases},
\end{eqnarray}
allowing us the implementation of all the local evolutions. 

Now, the implementation of the bilocal terms such as $e^{i\alpha Z_iZ_j}, e^{i\alpha Z_iG_j}$ and $ e^{\alpha Z_iY_j}$, where the last one corresponds to an interaction term in counterdiabatic approximation, where
\begin{eqnarray}
    Y_j = \begin{pmatrix}
        0& -i& 0\\
        i& 0 & -i\\
        0 & i & 0 
    \end{pmatrix},
\end{eqnarray}
can be done by using a controlled phase gate. In the case of trapped ions, it can be done through the center-of-mass motion, controlled by tuning a Raman transition to a specific red sideband as shown in Ref.~\cite{KilmovPhysRevA2003}. With an effective Hamiltonian, they model center-mass phonons by annihilation and creation operators, coupled to electronic transition $\ket{0}\rightarrow \ket{q}$, with $q =1,2$. In particular, we can obtain an effective unitary transformation between the qutrit $m$ and the center of mass of the form
\begin{eqnarray}
     &&U^{l,q}_m(\varphi) \ket{0}_m\ket{0} = \ket{0}_m\ket{0}\nonumber \\
    &&U^{l,q}_m(\varphi) \ket{0}_m\ket{1} = \cos\bigg(\frac{l \pi}{2}\bigg)\ket{0}_m\ket{1}-ie^{-i\varphi} \sin\bigg(\frac{l\pi}{2}\bigg)\ket{q}_m\ket{0}\nonumber\\
    &&U^{l,q}_m(\varphi) \ket{q}_m\ket{0} = \cos\bigg(\frac{l \pi}{2}\bigg)\ket{q}_m\ket{0}-ie^{i\varphi} \sin\bigg(\frac{l\pi}{2}\bigg)\ket{0}_m\ket{1}
\end{eqnarray}
where $l\pi/2 = \Omega_q \eta t/2$, with $\Omega_q$ is the effective Rabi frequency of transition $\ket{0}\rightarrow \ket{q}$ after adiabatic elimination of upper excited levels, $\eta $ is the Lamb-Dicke parameter and $\varphi$ is the laser phase. This interaction allows us to produce any controlled-phase gate through the center of mass, obtaining the basic key ingredients for universal qutrit quantum computing~\cite{WangFrontPhys2020}, and therefore an experimentally feasible way to implement our HDCQC paradigm. 

Related to the scalability of our approach, we can observe that in the fully digital implementation, in the worst case, each two-body term of the form $\ketbra{jk}{lm}$ can be implemented using one controlled phase gate and local gates, as in qubit-based quantum computing. As the error induced by the local gates is much less than the error of the two-body ones, the scaling of the error in our HDCQC paradigm is given by the number of controlled phase gates, which in the worst-case scale as the number of two-body terms in the Hamiltonian, similarly that in the qubit-based quantum computing paradigm, then different encodings do not provide enhancement

A more efficient implementation for current devices can be obtained by considering native interactions to implement HDCQC in the digital-analog paradigm~\cite{Kumar2024arXiv} or considering only algorithms in the impulse regime, where single-layer algorithms can be implemented as is shown in Ref.~\cite{Guan2024QuantumSciTechnol} for qubits beyond QUBO formulation.

\section{Data availability}
Data is available from the corresponding author upon reasonable request.

\section{Code availability}
The corresponding author will provide the code used for data analysis and plot generation upon reasonable request.

\section*{Acknowledgments}
We acknowledge financial support from Agencia Nacional de Investigaci\'on y Desarrollo (ANID): Financiamiento Basal para Centros Cient\'ificos y Tecnol\'ogicos de Excelencia grant No. AFB220001, Fondecyt Regular grant No. 1231172, Subvenci\'on a la Instalaci\'on en la Academia grant No. SA77210018. Also, the financial support from Vicerrector\'ia de Investigaci\'on, Innovaci\'on y Creaci\'on, USACH: JUV$\_$INVESTIGADORA$\_$DICYT$\_$USA21991 Grant No. 042431AA$\_$JUVI.

\section*{Author Contributions}
D. T. performs all the numerical and analytical calculations, as well as figures production. F. A.-A. provides the original idea, supervises the work, checks the calculations and writes the manuscript. Both authors agree with the final version of the work.

\section*{Competing interests}
The authors have no competing interests as defined by Nature Portfolio, or other interests that might be perceived to influence the results and/or discussion reported in this paper.

\end{document}